\documentstyle[preprint,aps,epsf]{revtex}

\tightenlines
\begin{document}
\draft

\title{
Structural properties of molten silicates from {\it ab  initio}
molecular-dynamics simulations: comparison
between CaO-Al$_2$O$_3$-SiO$_2$ and SiO$_2$.}

\author{ Magali Benoit and Simona Ispas}
\address{ Laboratoire des Verres, Universit\'e Montpellier II, \\ 
         Place E. Bataillon, 34095 Montpellier, France }
\author{ Mark E. Tuckerman}
\address{ Department of Chemistry and Courant 
Institute of Mathematical Sciences, \\
New York University, New York, NY 10003 USA 
}

\date{\today}
\maketitle

\begin{abstract}
We present the results of first-principles molecular-dynamics simulations of molten silicates,
based on the density functional formalism. 
In particular, the structural properties 
 of  a calcium aluminosilicate $ [$  CaO-Al$_2$O$_3$-SiO$_2$ $ ]$ melt are compared
to those of a silica melt. The local structures of the two melts are in
good agreement with the experimental understanding of these systems.
In the calcium aluminosilicate melt, the number of
non-bridging oxygens found is in excess of the number obtained from a simple
stoichiometric prediction.  In addition, the aluminum avoidance principle,
which states that links between AlO$_4$ tetrahedra are absent or rare,
is found to be violated. Defects such as 2-fold rings and 5-fold coordinated
silicon atoms  are found in comparable proportions
 in both liquids.
However, in the calcium aluminosilicate melt, 
a larger proportion of oxygen atoms are 3-fold coordinated.  In addition,
5-fold coordinated aluminum atoms are observed.
Finally evidence of creation and anihilation
of non-bridging oxygens is observed, with these oxygens being mostly
connected to Si tetrahedra.

\end{abstract}



\section{Introduction}
\label{sec:intro}

Silicate melts are the precursors to industrially and technologically 
important materials including ceramics
and nuclear and industrial waste confinement 
glasses.  They also occur naturally in the form of geologic
magmas.  Despite their technological relevance and geophysical 
importance, however, their
microscopic characteristics are not well understood~\cite{silicate}. 
The primary reason for this 
is that the structure of the melt is far more difficult to characterize
than that of crystalline silicates, necessitating a combination of indirect
methods.  Most of the information about the liquid structures come from
extrapolations of studies of glasses 
\cite{bond_length,feldspar,cormier,raman,cas_pressure,xanes,xray1,xray2,Al-coord4,calcul-Qn,stebbins_excess_nbo,flow},
but large changes in properties, such as heat capacity,
thermal expansion, compressibility, etc., as a result of heating suggest
significant structural changes with increasing temperature 
\cite{densite-liquide,sio-breaking,diffusion}.

In recent years, theoretical studies based on classical 
molecular-dynamics (MD) simulations have been able to treat binary or ternary silicate
systems with reasonable success.  However, accurate and reliable descriptions
 of systems containing more 
than three different atomic species by classical MD methods has proved far
more difficult, although several studies have been able to predict
general trends that are consistent with experimental 
results \cite{md-pression,viscosite-model,delaye-ghaleb,cormack,diffusion_Na}. 
This problem is particularly acute when cations such as Na$^+$ or Ca$^{2+}$
are introduced in silicate glasses, as it is
generally difficult to find interatomic potentials that can accurately treat
both the covalent and the ionic nature of 
the interactions and are also capable of describing the bond breaking and 
forming events that can occur in such chemically diverse environments.
These difficulties can be circumvented by employing the 
{\it ab initio} molecular dynamics method, in which 
internuclear interactions are calculated ``on the fly'' from
electronic structure calculations.  Indeed, recent {\it ab initio} MD studies 
of silica glass and of its melt have demonstrated 
the ability of this approach to describe the local structure and
dynamics of such systems with reasonable accuracy. These studies have also 
highlighted the advantages of employing a method which, additionally, 
allows direct access to the {\it electronic} 
properties of the system \cite{sio2_cp,EPJB}.

It is well known that many 
important macroscopic properties of silicate melts, such as the viscosity,
the glass transition temperature $T_g$, or the resistance to 
chemical change, for example
via corrosion, are dramatically altered by changes in the 
composition \cite{silicate,densite-liquide,viscosite-model}.
For instance, introduction of Na$^+$ ions into a SiO$_2$ melt causes
the viscosity at a given $T_g/T$ ratio to decrease by several
orders of magnitude from that of the  pure SiO$_2$ \cite{silicate}.
The presence of cations such as Na$^+$, K$^+$, Ca$^{2+}$ or  Mg$^{2+}$,
known as 
network modifier cations,  induces such changes by breaking some 
fraction of the Si-O bonds thereby creating non-bridging oxygens and disrupting
the tetrahedral silicate network.
Non-bridging oxygens (NBO) are oxygen atoms which do not connect two
tetrahedral cations or network-forming atoms, such as Si.
Non-bridging oxygens provide relatively weak connections between the
network forming atoms and the network modifier cations. 
However, when other network-forming atoms such as aluminum are introduced  into
the system, there is a gradual conversion of non-bridging oxygens into
bridging oxygens. This arises from the fact that most of the Al atoms
are tetrahedrally coordinated (AlO$_4^-$), and the resulting  negative charge
compensates the positive network modifier cation charge.
In such cases,
non-bridging oxygens can be created and the network broken 
only if there is an {\it excess} of network modifier cations, and it is for this 
reason that the viscosity of   ternary liquids, such as CaSiO$_3$
or Na$_2$SiO$_3$, progressively increases as the network modifier oxide (CaO or
Na$_2$O) is replaced by Al$_2$O$_3$.  The conventional explanation for
the increase in viscosity is the transformation 
of non-bridging oxygens into bridging oxygens as the concentration 
 of Al$_2$O$_3$ is increased.
Generally, the number of NBO can be predicted based on 
a knowledge of the composition by simple stoichiometric arguments.  
However, it  has  recently been shown that
such simple stoichiometric predictions are
not exactly fulfilled in calcium aluminosilicate (CAS) glasses and that a small
proportion of NBO can be present even if all the modifier cations
should, in principle, exactly compensate the AlO$_4^-$ tetrahedra 
\cite{stebbins_excess_nbo,flow}.

In this paper, the results
of an {\it ab initio} molecular-dynamics simulation of
a calcium aluminosilicate (CAS) melt are presented and its
microscopic characteristics are compared to those of a pure silica melt.
To our knowledge, these are the first fully {\it ab initio} MD studies of
the CAS melt.
We have chosen a CaO-Al$_2$O$_3$-SiO$_2$ system with a composition 
as close as possible to the basic composition of the confinement matrix for
the nuclear wastes.  This system also possesses a local structure
close to those of some rapid cooling magmas.
The chosen composition contains more Ca$^{2+}$ ions than are
needed to compensate the AlO$_4^-$ tetrahedra, thus leading to the
formation of non-bridging oxygens.
By carrying out a comparative study of the CAS and pure silica melts, it is
possible to describe the
detailed modifications in the network due to the presence
of Al and Ca$^{2+}$ in the system.

This paper is organized as follows:  In Sec.\ref{sec:methods}, 
the {\it ab initio} methodology is briefly described and the
details of the particular simulations performed here are
given.  In Sec.\ref{sec:results}, main results of the comparative study are
presented, including structural properties of the
CAS and silica systems. These results are discussed in Sec.\ref{sec:discussion}
 and conclusions are given  in Sec.\ref{sec:concl}.


\section{Simulation details}
\label{sec:methods}


Equilibrated configurations of the two liquids 
were generated by classical molecular dynamics runs, the details of which
are described in Secs. \ref{sec:sio2} and \ref{sec:cas} below, and were
subsequently used to initialize the {\it ab initio} MD simulations.
The two systems were then equilibrated within a Car-Parrinello (CP) 
{\it ab initio} MD run \cite{cp_85}
performed with the {\it ab initio} MD code, CPMD \cite{CPMD}.
In the {\it ab initio} MD simulations, the electronic structure was treated via
the Kohn-Sham (KS) formulation of 
density functional theory \cite{KS} within the local density approximation
for the pure silica system and within the generalized gradient approximation
for the CAS system employing the B-LYP functional \cite{becke88,lyp}.
The KS orbitals were expanded in a plane-wave basis at the $\Gamma$-point
of the supercell up to an energy cutoff of 70 Ry for both systems.
Core electrons were not treated explicitly but were replaced by
atomic pseudopotentials of the Bachelet-Hamann-Schl\"{u}ter type for
silicon \cite{BHS} and the Troullier-Martins type for oxygen \cite{MT}.
A Goedecker-type pseudopotential \cite{SG} was employed for aluminum,
and a Goedecker-type semi-core pseudopotential was employed for calcium.
The choices of the pseudopotentials, 
exchange and correlation functionals and plane-wave cutoff
are justified by previous studies
carried out on amorphous SiO$_2$ \cite{EPJB} as well as 
total energy calculations carried out on small molecules 
(see Table \protect\ref{table:tests}).


\subsection{Molten SiO$_2$}
\label{sec:sio2}

The molten silica system contains 26 SiO$_2$ 
units in a cubic box of edge length 10.558 \AA,
which corresponds to a mass density of 2.2 g$\cdot$cm$^{-3}$. The density of
the glass was chosen so that the configurations could later be used in quenching
runs to generate glass structures.
Although the density is, therefore, a little too high compared to the
real liquid, it is not expected that this will significantly affect our findings, which
are based on the comparison of network and disrupted network systems.
The SiO$_2$ initial configuration was obtained
by melting a 216 SiO$_2$ units $\beta$-cristobalite crystal at 7000 K with classical
molecular dynamics using 
the van Beest, Kramer and van Santen  (BKS) potential \cite{BKS,jund}
and then cooling it to 4200 K using the same potential.
At this temperature, a cubic box of edge length 10.558 \AA\ and containing
26 SiO$_2$ was extracted from the 216 SiO$_2$ system 
and equilibrated during $\sim$ 35 ps.

The classically equilibrated
SiO$_2$ liquid configuration 
was further equilibrated within a 6-ps {\it ab initio} MD run 
at 4200 K using a time step of 0.096 fs, then quenched to 3500 K at a
quench rate of 3 10$^{15}$ K$\cdot$s$^{-1}$ with the same time step, and finally 
equilibrated at 3500 K for 6 ps using a time step of 0.108 fs.
In order to achieve rapid equilibration and efficient
canonical sampling of the system, a separate
Nos\'{e}-Hoover chain thermostat \cite{nhc} was placed on each ionic
degree of freedom (known as ``massive'' thermostatting~\cite{tome})
and an additional Nos\'{e}-Hoover chain thermostat was placed
on the electronic degrees of freedom \cite{nhc,tuck_94}.
In all cases, a fictitious electronic ``mass'' parameter, $\mu$ (having units of
energy$\times$time$^2$) of 600 a.u. was employed.


\subsection{Molten calcium aluminosilicate}
\label{sec:cas}

The CAS system contains 22 SiO$_2$, 4 Al$_2$O$_3$ and 7 CaO, 
which gives approximately 67 $\%$, 12 $\%$ and 21 $\%$ molar percentages of
these units, respectively, and a total of 100 atoms.
For this particular composition, there are 8 Al atoms which
give rise to 8 AlO$_4^-$ tetrahedra under the assumption that
all Al atoms  form tetrahedra.  Four of the Ca$^{2+}$ ions then
compensate the negative charges of the AlO$_4^-$, leaving three Ca$^{2+}$ 
that can break the network and create, ideally, 6 non-bridging oxygens.
The system is confined in a cubic box with an edge length of  11.3616 \AA, 
which corresponds to a mass
density of 2.4 g$\cdot$cm$^{-3}$.
This density has been chosen by extrapolating to 2500 K the data obtained by Courtial
and Dingwell for a system  of close composition \cite{densite-liquide}.
For this case, classical MD simulations were also carried out on systems
containing 100 and 5184 atoms with the same composition described
above.  By comparing the structural properties obtained from the
classical MD simulations at the two system sizes, it was possible to estimate the 
finite-size effects on the {\it ab initio} MD data  and to 
validate the choice of  the system size for the {\it ab initio} simulations
\cite{mag-CEA}. 

The initial configuration of the CAS system for the {\it ab initio} MD simulation
was generated using a
Born-Mayer-Huggins potential \cite{delaye-ghaleb} in a classical MD run
to obtain a melt at 2000 K.
The CAS liquid was then heated to 2300 K with CPMD and further
equilibrated for 2 ps using a time step of 0.12 fs and
an electron mass parameter
of 800 a.u.  It was then heated again to 
3000 K and equilibrated for 6.8 ps, during which stronger diffusion effects 
occurred than at the lower temperature. 
Again, rapid equilibration and
efficient canonical sampling was
achieved by coupling a Nos\'{e}-Hoover chain thermostat \cite{nhc,tuck_94}
to each ionic degree of freedom.


\section{Presentation of Results}
\label{sec:results}


In this section, the structural properties of the CAS and SiO$_2$
melts are presented in terms of network pair correlation
functions, angle distributions, examination of Al-O-Al linkages,
proportion of NBO, and cation pair correlation functions.  
A full discussion of these results is presented in Sec.~\ref{sec:discussion}.

\subsection{Network pair correlation functions}
\label{sec:network}

In this subsection, the pair correlation functions (PCF) 
corresponding to the network-forming atoms (Si, O and Al for CAS) are
presented for the CAS melt and compared to those of the silica melt
when appropriate.

\begin{figure}
\epsfxsize=300pt{\epsffile{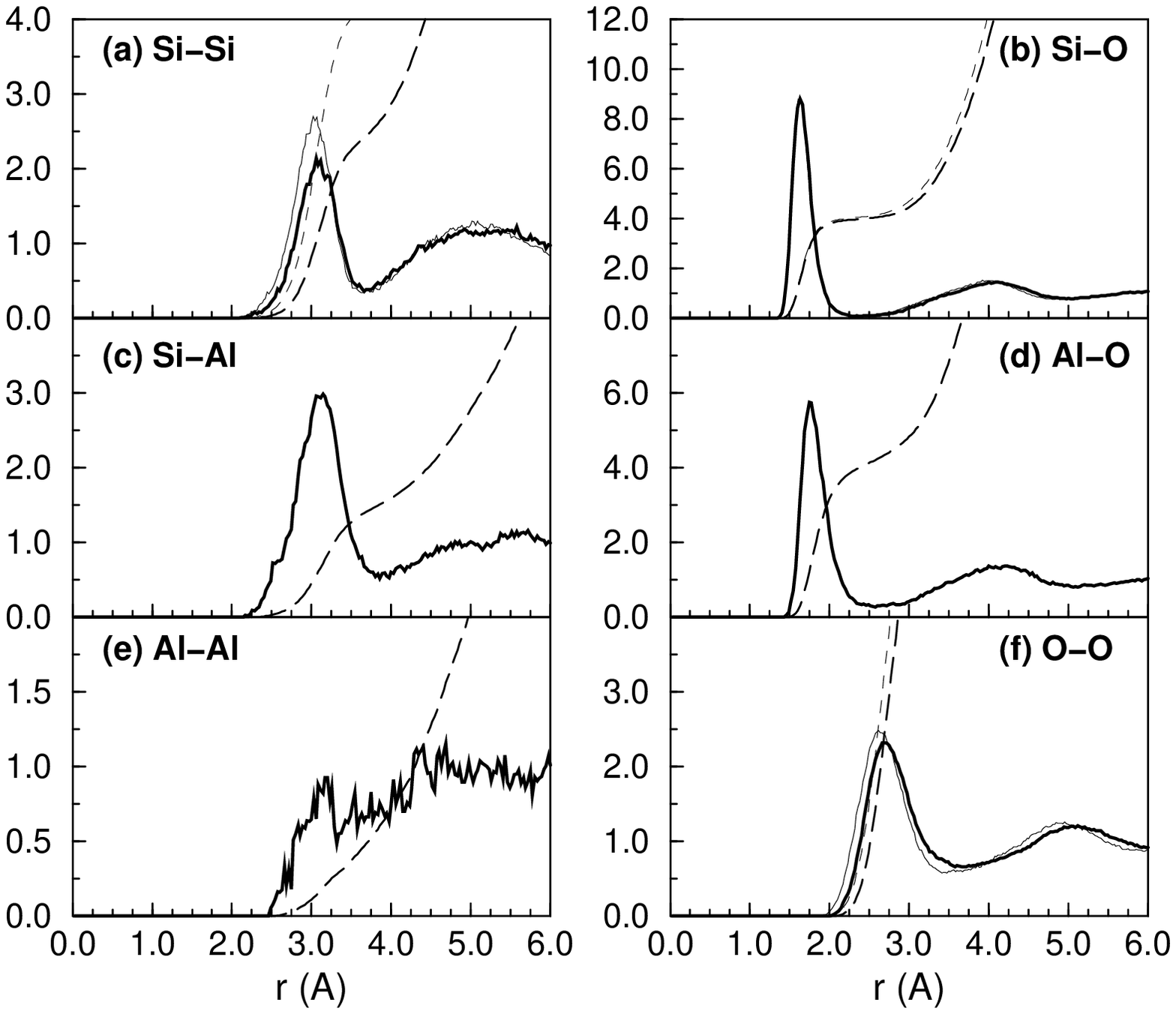}}
\caption{
Pair correlation functions of the network-forming atoms in the CAS system 
(bold lines) and in the silica system (thin lines). The integrated coordination
numbers are also shown in dashed lines (bold and thin for CAS and SiO$_2$, respectively).}
\label{fig:GOFR_NET}
\end{figure}

The Si-O PCF and the corresponding integrated coordination
number are almost identical for the two systems, which confirms that the basic 
tetrahedral unit is conserved between these two liquids 
(Fig. \protect\ref{fig:GOFR_NET}(b)). 
Moreover, the similarity in CAS between the Al-O (Fig. \protect\ref{fig:GOFR_NET}(d))
and Si-O PCFs is clear evidence of the fact that the Al atoms can substitute for
the Si atoms at the center of the tetrahedra.  The ability of Al to replace Si in the
network has also been observed in experimental studies of aluminosilicate
glasses \cite{xanes,xray1,xray2,Al-coord4,calcul-Qn}.
The slight shift of the Al-O peak to a higher $r$ value is consistent with
previously observed and calculated Al-O and Si-O bond lengths in 
aluminosilicates \cite{bond_length} as well as with the larger covalent radius of Al
compared to Si.  

Comparison of the Si-Si and O-O PCFs between the CAS and SiO$_2$ liquids also shows
a slight shift of the first peak toward higher $r$ values in the CAS case, and the
plateau in the running Si-Si coordination number is considerably lower in 
the CAS system.
In order to explain the shift in the Si-Si peak, we first note that
in silica, the first Si-Si neighbors correspond to neighboring tetrahedra
(i.e. the first Si-Si distances correspond to the Si-O-Si linkages).
In the CAS system, however, some of the oxygens, in particular, the NBO, 
are connected to only one network forming atom (Si or Al) and one
network modifier (Ca), thus forming
Si-O-Ca or Al-O-Ca linkages.  Therefore, some of the first-neighbor Si-Si
distances are due to these more complex linkages.  
The reduction in the Si-Si coordination is due primarily 
to the fact that some of the Si atoms
are replaced by Al atoms. 
As a result of the Al substitution for Si {\it and} the presence of NBOs,
there are fewer direct Si-O-Si linkages in CAS, leading to an average
 coordination of 2.4 compared to 4 in the pure silica system.
The shift in the position of the first O-O peak
in the CAS system is simply a reflection of the fact that the Al-O bond length is
larger than that of Si-O.  Thus, if Al substitutes for Si, maintaining both
the regular tetrahedral coordination and the angles between neighboring tetrahedra
as in silica, then the observed shift in the O-O peak is expected.

\subsection{Angles distributions}
\label{sec:angles}

Evaluation of the angles distributions and coordination
numbers presented in this subsection was based on distance cutoffs
determined from the 
first minimum of the PCFs (2.38 \AA\ for  Si-O and 2.56 \AA\ 
for Al-O).

\begin{figure}
\epsfxsize=280pt{\epsffile{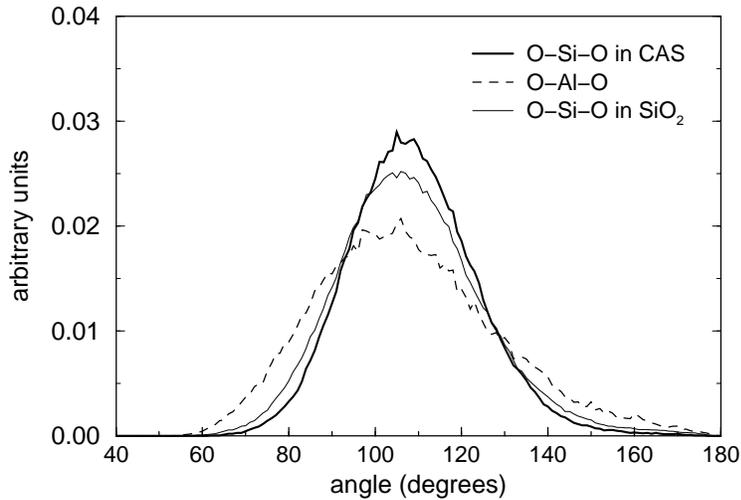}}
\caption{
Distribution of the O-Al-O angles in CAS (dashed line) and 
of the O-Si-O angles in CAS (bold line) and in SiO$_2$ (thin line).}
\label{fig:Angles_tetra}
\end{figure}

In order to investigate the effect of Al substitution for Si 
on the tetrahedral angles,
the distributions of the O-Al-O
and the O-Si-O angles for both systems were computed and are 
shown in Fig. \protect\ref{fig:Angles_tetra}.
The figure shows that
the O-Al-O distribution in CAS 
is somewhat broader than the O-Si-O distribution in both CAS and in silica.
The mean values of these distributions, however, are very similar 
(108.8 $^{\circ} \pm 15$ for O-Si-O in
CAS, 108.3 $^{\circ} \pm 17$ for O-Si-O in SiO$_2$ 
and 107.6 $^{\circ} \pm 21$ for O-Al-O), suggesting
that the angles between tetrahedra remain approximately unchanged 
upon Al substitution.
Therefore, the primary cause of the shift in the first peak of the O-O PCF 
is the increased Al-O bond length.

\begin{figure}
\epsfxsize=340pt{\epsffile{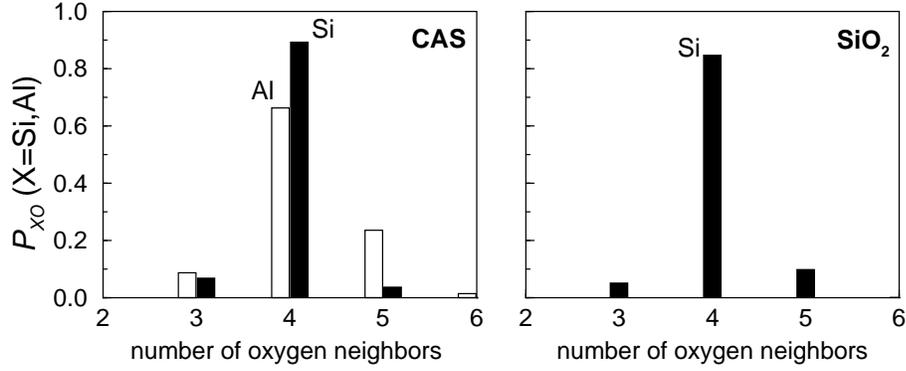}}
\caption{%
Left graph: Histograms of $P_{XO}(N) \  (X=Si,Al)$, the probability of finding 
$N$ oxygen atoms around the silicium (black) and aluminum (white) atoms in CAS.
Right graph: Histogram of $P_{SiO}(N)$, the probability of finding 
$N$ oxygen atoms around the silicium atoms in SiO$_2$.}
\label{fig:COORD_network}
\end{figure}

The relative broadness of the O-Al-O angle distribution compared to that of
O-Si-O likely signifies an Al-O coordination that is different from 4.
In order to check this assertion, the average oxygen coordination of the
Al and Si atoms was computed.
  The histogram of the number of oxygen neighbors
of Al and Si is shown in Fig. \protect\ref{fig:COORD_network} (left panel). 
From the figure, it can be seen that, although Al substitutes for Si in 
the network, the fraction of 3-fold and 5-fold coordinated Al exceeds 
that of Si.  
Recent high-temperature NMR measurements of aluminosilicate 
melts~\cite{diffusion,MgAlSiO} directly revealed the presence of
6-fold coordinated Al and strongly suggested the possibility of
5-fold coordinated Al, although the latter were not directly 
observed in these experiments.
However, MAS NMR experiments~\cite{alo5,new_al5} 
have provided clear evidence of the existence of small amounts of
both 5- and 6-fold Al coordination 
in binary Al$_2$O$_3$-SiO$_2$ and in ternary 
CaO-Al$_2$O$_3$-SiO$_2$ glasses \cite{alo5,new_al5}. 
  Indeed, the average proportion of 5-fold coordinated Al 
in the present simulations is also small compared to 4-fold coordinated Al.
However, we  observe less than 1 $\%$ of 6-fold coordinated Al for this particular composition.

\begin{figure}
\epsfxsize=280pt{\epsffile{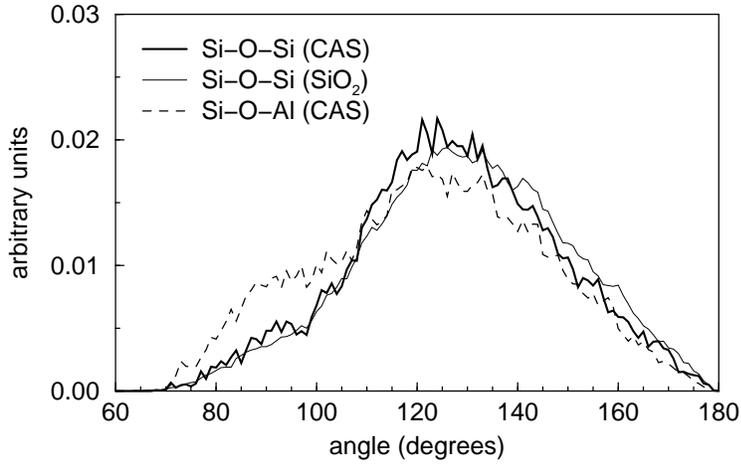}}
\caption{%
Distribution of the Si-O-Si angles in  CAS (bold line) and in SiO$_2$ (thin line)
and distribution of the Si-O-Al angles in CAS (dashed line).}
\label{fig:Angles_SiOSi}
\end{figure}

We also investigated the effect of the Al substitution for Si
on the angles between neighboring tetrahedra. The Si-O-Al and 
Si-O-Si angles distributions  were computed for 
the CAS melt and compared to 
the Si-O-Si angle distribution for the SiO$_2$ melt 
(Fig. \protect\ref{fig:Angles_SiOSi}). 
The Si-O-Si distributions are very similar in the two systems, 
only showing a slight difference in the intensity of the shoulder 
around 90 $^{\circ}$- 100 $^{\circ}$.
 It can be shown that values of the Si-O-Si angles 
around 90 $^{\circ}$- 100 $^{\circ}$ can be attributed to 2-membered rings 
and/or to oxygen tri-clusters, 
oxygens bound to three network-forming atoms (see Sec.~\ref{sec:discussion}).
A comparison of the proportions of these units in the two different systems 
is consistent with the differences observed in the angle distributions and
will be  discussed in more detail in section \protect\ref{sec:discussion}.
In CAS, the Si-O-Al angle distribution shows a higher shoulder around 90 $^{\circ}$- 100 $^{\circ}$
than the Si-O-Si one. Again the Si-Al 2-membered rings and  the oxygen tri-clusters
 are at the origin of this shoulder. 
Overall, the introduction of Al and Ca does not significantly affect
the Si-O-Si angle distribution in the molten state.

\subsection{Al-O-Al linkages}
\label{sec:aloal}

Although the Al-Al PCF (Fig. \protect\ref{fig:GOFR_NET}(e)) 
is somewhat noisy due to the small number of Al atoms in the system, 
this correlation function exhibits a first peak at $\sim$ 3.2 \AA, which indicates that 
some Al-O-Al linkages are present in the system.  This would appear to be in 
direct violation of the so called
Al avoidance principle or L\"{o}wenstein's rule~\cite{model}, 
an empirical rule which states that two
Al tetrahdera are never found linked by an oxygen atom in aluminosilicate
crystals and glasses, at low Al content.
However this rule has been found experimentally not to be exactly
fulfilled in some glasses and melts,
{\it in particular} when Ca atoms are present,
leading to Al/Si alternance disorder~\cite{al-avoidance1,al-avoidance2,himmel}.

The degree of Al avoidance violation can be quantified by
examining the average number of Al-O-Al linkages formed over the course of the
simulation.  
By computing the number of Al atoms around each oxygen atom 
(using the first minimum of the $g(r)$, 2.56 \AA, as the Al-O cutoff distance)
and counting one Al-O-Al linkage when two Al atoms are found, an average value of 
2.26 Al-O-Al linkages is obtained, i.e., $\approx$ 57 $\%$ 
of Al atoms form Al-O-Al linkages, if they do not form chains. Moreover, 
half of
the oxygen atoms involved in the Al-O-Al linkages are found to be 3-fold coordinated, on average. 
The proportion of Al-O-Al linkages is greater than that which would be obtained 
from a purely random model~\cite{al-avoidance2}. 
Thus, the Al atoms appear to favor Al-O-Al linkages in the CAS liquid
and adhere only minimally to the Al avoidance principle.  That the
presence of Ca$^{2+}$ cations preferentially favor Al-O-Al linkages and, thus,
strong violation of the Al avoidance principle, has also been suggested by
static {\it ab initio} calculations of clusters~\cite{ab-initio}
and is likely due to the greater aggregation of negative charge
around the Al-O-Al linkage.  In a disordered (liquid or glass) state, 
the relatively large Ca$^{2+}$ charge is, therefore, able to 
induce formation of such linkages.

\subsection{Non-bridging oxygens}
\label{sec:NBO}

\begin{figure}
\epsfxsize=280pt{\epsffile{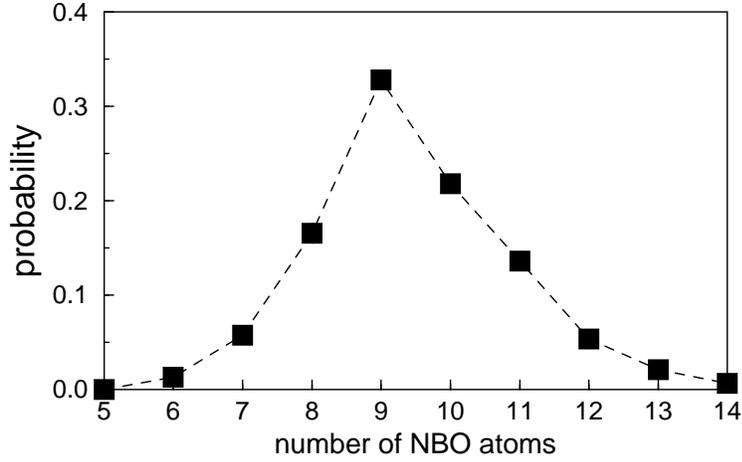}}
\caption{%
Average distribution 
of the number of NBO atoms in the CAS system.
The size of the symbols indicate the maximum error bar associated
with the choice of distance cutoff for Al-O and Ca-O.}
\label{fig:NBO}
\end{figure}

In the present CAS system, stoichiometry dictates that four of the 
Ca$^{2+}$ ions should compensate the eight aluminum tetrahedra (assuming all the
Al are four-fold coordinated), leaving three Ca$^{2+}$ cations to create
six non-bridging oxygens. The number of NBOs in the system can be computed
by counting the number of oxygen atoms which have only one Si or Al 
neighbor and one Ca neighbor. The neighbors are defined to be any atoms
within a sphere of radius determined by the first minimum of the corresponding 
PCF (2.38 \AA\ for O-Si, 2.56 \AA\ for O-Al, and 3.40 \AA\ for O-Ca --
see Sec. \protect\ref{sec:network} and Sec. \protect\ref{sec:calcium}),
centered on each oxygen atom.
The distribution of the number of  non-bridging oxygens found in the CAS system
during the simulation is depicted in Figure~\protect\ref{fig:NBO}. 
The distribution is peaked at 9, while the probability of the 
system's possessing six NBOs is relatively small.  Thus, the average number 
of NBOs in the system is larger than would be predicted from a simple 
stoichiometric argument.  
In a recent experiment of Stebbins and Xu~\cite{stebbins_excess_nbo}, it was 
found that in a particular CAS glass system in which the Ca$^{2+}$ ions perfectly
compensate all of the negative Al tetrahdera, some NBOs are present, although
stoichiometric arguments would predict that there should be none.
Since the present study is concerned with the liquid state, where bond-length
fluctuations and other dynamical effects are more significant than in the
glass, we cannot directly compare our result with that of
Stebbins and Xu.  However, given that the trend is toward a {\it larger} number
of NBOs in the glass than the stoichiometry would predict, 
the fact that our calculations predict a similar trend in the liquid accords well
with the experimental result.

\begin{figure}
\epsfxsize=320pt{\epsffile{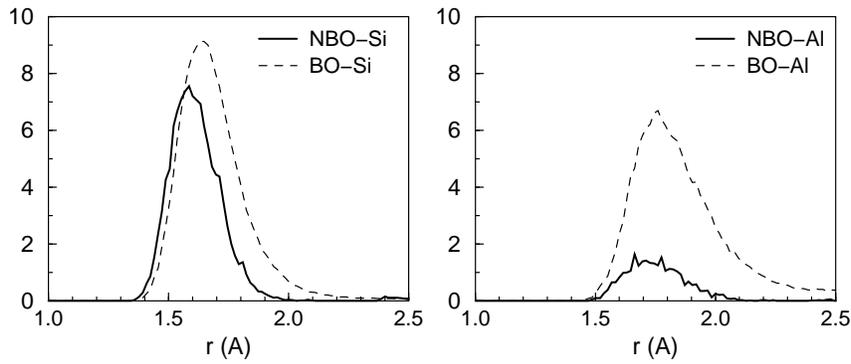}}
\caption{%
Left graph: First peaks of the Si-BO (dashed line) and Si-NBO (bold line) pair correlation
functions in CAS.
Right graph: First peaks of the Al-BO (dashed line) and Al-NBO (bold line) pair correlation
functions in CAS.}
\label{fig:GOFR_NBO}
\end{figure}

In  Figure \protect\ref{fig:GOFR_NBO}, we present the first peaks of 
the PCFs between the oxygen atoms and the network-forming atoms (Si and Al),
evaluated  separately for bridging (BO) and non-bridging (NBO) oxygen atoms.
The maximum intensity of the X-BO and X-NBO  (X=Si,Al) peaks are located at
different $r$ values, the X-NBO distances being shorter than the corresponding
to those for X-BO. This result is in agreement with experimental results concerning  
sodium silicates and aluminosilicates \cite{bond_length,si-nbo-bond} and with 
 {\it ab initio} calculations on  clusters 
\cite{ab-initio,ab-initio-ns4}. In the simulated CAS melt, 
the most probable distances are $r({\rm Si-BO})$ $\approx$ 1.64 \AA\ and $r({\rm Si-NBO})$
$\approx$ 1.58 \AA, $r({\rm Al-BO})$ $\approx$  1.75 \AA, and $r({\rm Al-NBO})$
$\approx$ 1.70 \AA.  It is also interesting to note that the  Si-O and Al-O distances 
in the molten state are very close to their respective values in the glass.

\subsection{Calcium pair correlation functions}
\label{sec:calcium}

In this subsection, structural properties involving the
calcium atoms in the CAS melt, 
such as PCFs and coordination numbers, are presented.

\begin{figure}
\epsfxsize=300pt{\epsffile{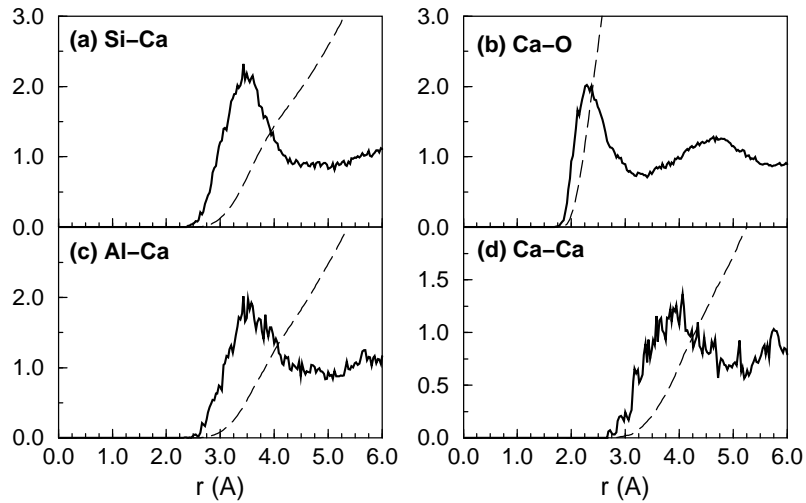}}
\caption{%
Pair correlation functions of the calcium atoms in the CAS system
and the corresponding integrated coordination numbers.}
\label{fig:GOFR_Ca}
\end{figure}

\begin{figure}
\epsfxsize=280pt{\epsffile{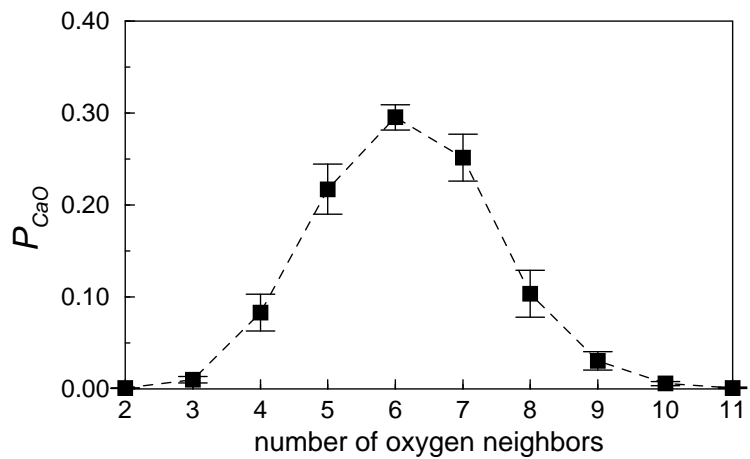}}
\caption{%
Probability distribution, $P_{CaO} (N)$, for finding $N$ 
oxygen atoms around a calcium atom in the CAS system. 
The error bars were estimated by bracketing the first
minimum of $g_{Ca-O}(r)$ between statistical lower
and upper bounds and varying the cutoff distance
between these bounds.}
\label{fig:COORD_Ca}
\end{figure}

The calcium PCFs are depicted in Fig. \protect\ref{fig:GOFR_Ca}.
The Ca-O radial distribution function (Fig.~\protect\ref{fig:GOFR_Ca}(b))
exhibits a first peak at approximately 2.33 \AA, in agreement with
experimental values obtained for glasses and minerals of similar composition
~\cite{feldspar,xray2}.
The coordination of calcium atoms is found to be 
 equal to 6.2 $\pm$ 1.3 
on average (see Fig. \protect\ref{fig:COORD_Ca}), while
the experimental value in 
calcium aluminosilicate {\it glasses} 
is estimated to be between 5 and 6~\cite{feldspar,xray2}.
This comparison is only meant as a qualitative one, since direct comparison between
the liquid and glass states is not possible.  Indeed, 
the high temperature of the molten state gives
rise to large fluctuations in the Ca-O coordination number.

The Ca-Si and Ca-Al PCFs ((Fig.\protect\ref{fig:GOFR_Ca}(a) and (c))
also show well defined first peaks at $\sim$ 3.45 \AA, which are due either to direct
bonds with non-bridging oxygens, i.e., Si-O-Ca and Al-O-Ca linkages, 
or to  the proximity of Ca$^{2+}$ ions to compensate the negative AlO$_4^-$ groups.
It can be shown that the Ca-Si peak is due to the former
and the Ca-Al to the latter. Indeed,
we observe mostly Si-NBO bonds and very few  Al-NBO bonds 
($\sim$ 91 $\%$ of the NBOs are connected to Si atoms), 
a result that is consistent with recent X-ray experiments on glasses of similar
composition \cite{xray2}. In Fig. \protect\ref{fig:GOFR_NBO},
in which PCFs are evaluated separately for bridging (BO) and
non-bridging (NBO) oxygen atoms, we observe that Si PCFs possess
similar characteristics for both BO and NBO.
In contrast, 
the Al-NBO PCF is generally small compared to the Al-BO PCF, which
indicates that almost all the NBO atoms are connected to Si tetrahedra.
Thus, the first peak in the Ca-Al PCF is due to 
nearby Ca$^{2+}$ ions which, in this close proximity, are able to 
compensate the negative AlO$_4^-$ groups.
Note that the Ca-Ca PCF, shown in Fig.~\protect\ref{fig:GOFR_Ca}(d)
does not exhibit a well defined first peak, which suggests that
at liquid conditions, there is no ordering of the network modifier cations.

\section{Discussion}
\label{sec:discussion}

In order to explain the presence of excess non-bridging oxygens,
Stebbins and Xu \cite{stebbins_excess_nbo} proposed two possible 
structural units: AlO$_5$ groups and tri-clusters.
Tri-clusters are oxygen atoms bonded to three network-forming
atoms, either Al or Si, and they can be of four types : oxygens bonded to three 
silicon atoms (3 Si), to
two silicon and one aluminum atoms (2 Si - 1 Al), to one silicon and two 
aluminum atoms (1 Si - 2 Al), and  to three aluminum atoms (3 Al). 
In the molten state we observed that the fraction of 3-fold coordinated oxygen atoms
is not negligible.  A more detailed analysis of these tri-clusters in CAS showed that, 
due to the small number of aluminum atoms in the system, 
 3-Al tri-clusters are absent and  that the most numerous tri-clusters
are the 2 Si - 1 Al and  2 Al - 1 Si. 
When added together and averaged over the trajectory,
the percentage of oxygen atoms forming tri-clusters is around 6.9 $\%$ 
in the CAS  system, with large fluctuations of about $\pm$ 2.8 $\%$. 
This result accords well with
the idea of  Stebbins and Xu \cite{stebbins_excess_nbo} 
that a given number of oxygen tri-clusters should be present 
in the glass in order to compensate the formation of the non-bridging oxygens. 
It could be argued that the relatively small number of tri-clusters may
be a consequence of thermal fluctuations at liquid conditions.  Although
thermal induced tri-cluster formation cannot be ruled out, it is worth
noting that  only 4.7 $\%$ of tri-clusters could be identified in the silica melt
at higher temperature.

On the other hand, in CAS, we also find a relatively 
high number of AlO$_5$ units:  on average, approximately $\sim$ 1.9
or 23\% of Al are 5-fold coordinated, however the fluctuations
are such that in any configuration, anywhere between 0 and 3 AlO$_5$ units may exist.
This result suggests that the presence of highly coordinated
aluminum atoms could favor the creation of excess NBO atoms. 
Most experimental studies \cite{xray1,xray2,Al-coord4,calcul-Qn}
only show evidence of 4-fold coordinated
aluminium atoms in aluminosilicate glasses. However, 
as discussed in Sec.\protect\ref{sec:angles}, experimental
evidence of higher Al coordination numbers (mainly 5 and 6-fold) in the
glass and  molten state of calcium aluminosilicate systems exists,
and, recently, 
evidence of high Al coordination in magnesium aluminosilicate glasses has
been reported \cite{densite-liquide,diffusion,MgAlSiO,alo5,new_al5}.

In Fig. \protect\ref{fig:NBO}, we present the 
distribution of the number of NBOs found during the simulation of the CAS melt.
The relatively large width of the distribution
indicates that the identity of the oxygen atoms (BO or NBO)
does  not remain constant during the simulation. 
Indeed, bond-breaking events that lead to the 
creation and the anihilation of non-bridging oxygens, 
i.e. Q$_4$ and Q$_3$ exchanges, are observed.
This mechanism has been suggested 
as underlying the induction of shear
flow and, hence, 
a decrease in the viscosity in these systems 
\cite{densite-liquide,sio-breaking,diffusion}.
Since  almost no NBOs are found to be connected
to Al atoms on average during the simulation (see Sec. \protect\ref{sec:calcium}), 
it is highly  probable that more Al-O bonds are broken than Si-O bonds.  
In silica, however, since there are no NBOs, a different mechanism
must underly the flow process.
Unfortunately, we can not extract direct dynamical quantities 
from the present simulations.
This is mainly due to the fact that at high temperature in the molten state,
the electronic gap is too small compared to $k_{\rm B}T$
to ensure the decoupling of the ionic and the electronic 
degrees of freedom, 
which is needed for {\it ab initio} molecular dynamics in a constant
energy ensemble.  
The use of thermostats 
becomes compulsory and the direct access to dynamical properties is no longer available.  
We are currently developing new techniques 
to treat this problem~\cite{Minary1}.

Nevertheless, the structural characteristics of the melt can  give some insight into its
dynamical properties. For instance, it has been suggested that 
five-coordinated network cations could act as transition states 
for the flow process \cite{densite-liquide,sio-breaking,diffusion,new_al5}. 
This atomic-scale flow step was described in Ref. \cite{flow} as an oxygen 
atom changing from bridging to non-bridging, 
or vice versa: ``A non-bridging oxygen 
 bonded to a modifier cation can approach a silicon atom and make it
over-coordinated. Dissociation of the over-coordinated SiO$_5$ results
in an exchange of the roles of oxygen from bridging to non-bridging ...''.
In the present simulations, we find a significant 
fraction of 5-fold coordinated 
silicon atoms ($\approx$ 9.9 $\%$ in SiO$_2$ at 3500 K and
$\approx$ 3.8 $\%$ in CAS at 3000 K) 
and a large fraction of 5-fold coordinated 
aluminum atoms in CAS (see Fig.\protect\ref{fig:COORD_network}).
These numbers show very large fluctuations around their average values which
supports the idea that these units participate in the transition state of 
the flow steps. 
The relatively large number of 5-fold coordinated 
network cations in CAS compared to SiO$_2$, even
at a lower temperature, could be a signature of a lower viscosity in the
aluminosilicate melt. It is also 
interesting to note that the number of 5-fold coordinated Si atoms
is larger in SiO$_2$ than in CAS. 
In the latter system, the 5-fold coordinated Al atoms, 
which are energetically more favorable, replace the 5-fold Si atoms 
as the transition state of the flow step.

\begin{figure}
\epsfxsize=280pt{\epsffile{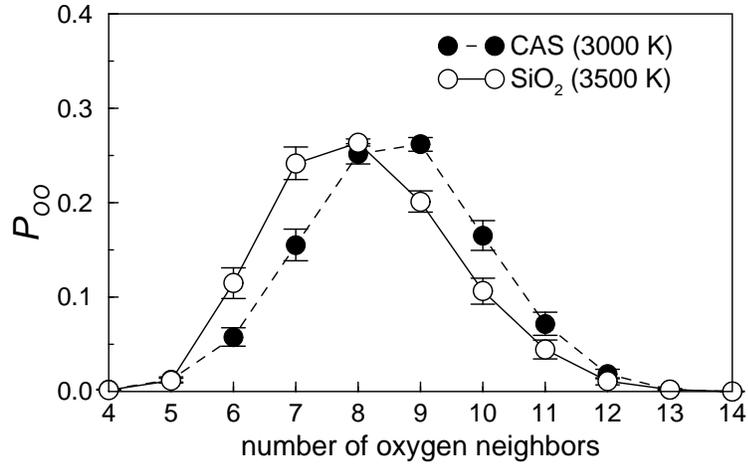}}
\caption{%
Probability distribution, $P_{OO}(N)$, for finding $N$ oxygens
around a given oxygen atom in CAS and SiO$_2$. 
The error bars were estimated by bracketing the first
minimum of  $g_{O-O}(r)$ between statistical lower
and upper bounds and varying the cutoff distance
between these bounds.}
\label{fig:Z_OO}
\end{figure}

In Sec.~\ref{sec:results}, it was seen that the local
order in the two silicate melts is only slightly affected by temperature and
by the introduction of a small number of modifier cations. 
The Si tetrahedral units remain the most probable basic units of the system, 
with probabilities of 85 $\%$ in SiO$_2$ and 89\% CAS at 3000 K.
From structural characteristics such as pair correlation functions or 
angle distributions, therefore, it is almost impossible to discern the disruption
of the silicate network by the modifier cations, at high temperature.
The effect can be seen, however, by looking at characteristic patterns in
the network. 
An example is the oxygen-oxygen coordination number, which is a more
sensitive probe of the network disruption.
We have evaluated these coordination numbers by counting the number 
of oxygen neighbors around each oxygen
atom inside a sphere of radius determined by the first minimum 
of the O-O radial distribution function. 
Histograms of the oxygen-oxygen coordinations 
are depicted in Fig. \protect\ref{fig:Z_OO} for
the two different systems. For SiO$_2$ at 3500 K, the distribution of 
the O coordination peaks around 7 or 8 atoms, which indicates that the oxygen 
atoms 
have more oxygen neighbors than would be expected from two connected tetrahedra.
In the SiO$_2$ glass, in which the network is not broken, the maximum 
of the distribution occurs at 6 which corresponds to the number 
of oxygen neighbors belonging to two connected tetrahedra. 
The fact that, at 3500 K, the maximum is located between 7 and 8 in silica shows
that part of the network has been broken. 
One of our hypotheses is that 2-membered rings and highly coordinated
network formers are the cause of the increased number of oxygen neighbors. 
Indeed, the percentage of 2-membered
rings in silica is not negligible at 3500 K ($\sim$ 12.4 $\%$).
It could be argued that our liquid silica system is too compressed  (its density
is equal to the experimental density of the silica glass at room temperature) and that
the high pressure induces the shift in the O-O correlation and the large number of 
edge-sharing tetrahedra. However, results of classical-molecular dynamics 
on liquid silica at several densities and temperatures show that this is not the case
\cite{vollmayr_phd,kob-horbach}. At $\approx$ 3200 K and zero pressure, the O-O distribution
ressembles the distribution in Fig. \protect\ref{fig:Z_OO} for silica \cite{vollmayr_phd} 
whereas at higher pressure (for a density of 2.35 g.cm$^{-3}$) and temperatures up to 4200 K, the
O-O distribution is peaked between 6 and 7, which is close to that of the
glass structure \cite{kob-horbach}. 
The shift of the O-O distribution with pressure 
is accompanied by a decreasing number of edge-sharing tetrahedra. 
This ``non intuitive'' behavior has already been seen in previous Monte Carlo simulations 
of liquid silica under pressure, in which it has been shown that the proportion of small rings
decreases with increasing pressure \cite{stixrude}.  
Given the similarities between our result and that of Ref. \cite{vollmayr_phd}, and given the
small thermal expansion of silica (0.54 $10^{-6}$ K$^{-1}$ at 1000 $^{\circ}$ C \cite{doremus}), 
it is clear that the high O-O correlation is a consequence of the high temperature of the system. 

The distribution of O-O coordination in CAS at 3000 K is shifted 
towards higher numbers
compared to that of SiO$_2$ (Fig. \protect\ref{fig:Z_OO}). 
In the former, the disruption of the network is not only due to the NBOs but 
also to the presence of highly coordinated Al atoms which weaken the structure
\cite{viscosite-model}. The CAS melt density is only 
slightly larger than that given by the extrapolation to 3000 K 
of experimental data \cite{densite-liquide}, 
therefore the high correlation of the oxygen atoms is clearly
due only to the high temperature.
The relative shifts of the O-O distributions between the two  different
systems can be related to the disruption of the three dimensional network.
In the two melts, thermal effects create defects such as 2-membered
rings and SiO$_5$ units in comparable proportions. In CAS, however, 
 the disruption of the network is also driven by the presence of 
non-bridging oxygens, which break some of the Si-O bonds, as well as by the presence of 
the aluminum atoms which possess high coordination numbers.
Indeed, the presence of aluminum atoms has already been proven to be
responsible for the increased fragility of the silicate network~\cite{viscosite-model},
which can be traced to the fact that the Al have  broader
coordination distributions, 
leading, therefore, to an increased flexibility of the network structure than in SiO$_2$.

The use of the Nos\'{e}-Hoover chain thermostatting method~\cite{nhc} 
in the present simulations allows the heat capacity at constant volume,
$C_v$ to be computed efficiently using the relation:
\begin{equation}
\Delta E = \sqrt{k_B T^2 C_v} 
\label{eq:cv}
\end{equation}
where $\Delta E = \sqrt{\langle E^2\rangle - \langle E\rangle^2}$ 
denotes the
fluctuations of the potential energy plus the 
kinetic energy of the ions over the course of
the simulation. In silica at 3500 K, we find that the heat capacity at constant volume
is equal to 91.5 J mol$^{-1}$ K$^{-1}$. As a first approximation, we can roughly estimate the
difference between C$_v$ and  C$_p$, the heat capacity at constant pressure, which is the
experimentally measured quantity:
\begin{equation}
C_p - C_v = T V \frac{\alpha^2}{\beta} 
\label{eq:diff}
\end{equation}
where $\alpha$ is the thermal expansion coefficient and $\beta$ the isothermal compressibility. 
Using the experimental values $\alpha$=0.54 10$^{-6}$ K$^{-1}$, valid at 1000 $^{\circ}$ C, and
$\beta$ = 2.74 $10^{-2}$ GPa$^{-1}$ at 300 K \cite{doremus}, we find a difference of 
C$_p$ - C$_v$ = 1.0 10$^{-3}$ J mol$^{-1}$ K$^{-1}$, so C$_p$ $\approx$ 91.5 J mol$^{-1}$ K$^{-1}$.
This value does not disagree with an extrapolation of the experimental values of C$_p$ from
Ref. \cite{Cp_sio2} and is close to the values obtained by Scheidler {\it et al.} \cite{kob_cp}
from molecular dynamics simulation of the SiO$_2$ melt in the same temperature range, 
using the BKS potential \cite{BKS}. Scheidler {\it et al.} computed the 
frequency dependent specific heat from $T$=2750 K 
up to $T$=6100 K and found a reasonable extrapolation
of the experimental $C_p(T)$ above $T_g$.

For the CAS melt at 3000 K, we find a value of C$_v$ = 99.28 J mol$^{-1}$ K$^{-1}$ and
an estimate of the difference, C$_p$ - C$_v$ = 0.189 J mol$^{-1}$ K$^{-1}$, 
which gives 
C$_p$ $\approx$ 99.47 J mol$^{-1}$ K$^{-1}$. The C$_p$ - C$_v$ 
difference has been estimated using experimental values for calcium aluminosilicate systems
of close compositions: $\alpha$ = 54.5 10$^{-7}$ K$^{-1}$ at 815 $^{\circ}$ C (25 mol$\%$ CaO, 15 mol$\%$ Al$_2$O$_3$, 60 mol$\%$ SiO$_2$)  \cite{Mazurin}
and $\beta$ = 1.28 $10^{-2}$ GPa$^{-1}$ at 300 K 
(26.7 mol$\%$ CaO, 13.3 mol$\%$ Al$_2$O$_3$, 60 mol$\%$ SiO$_2$)  \cite{doremus}. 
Although this estimate of C$_p$ is relatively crude,
it can be used to give an order of magnitude for C$_p$ at high temperature. 
To our knowledge, no experimental values of C$_p$ 
have been reported for a CAS system 
of close composition,  at high temperature.


\section{Conclusion}
\label{sec:concl}

The structural properties of two different molten silicates have been analyzed
using first-principles molecular-dynamics.
It has been observed that, even in the calcium aluminosilicate melt, 
the basic structural unit of the system, i.e. the SiO$_4$ tetrahedron
is not destroyed at high temperature and that the aluminum
atoms can substitute for the silicon atoms in the center
of the tetrahedra.
Analysis of the structure of two melts show
that the temperature effects induce the creation 
of 2-membered rings, 3-fold coordinated oxygen atoms
 and 5-fold cooridnated silicon and aluminum atoms. 
In CAS, in particular, a relatively large proportion of
Al atoms are found to be 5-fold coordinated.

In the CAS system, a larger number of non-bridging oxygens than would
be predicted from stoichiometry was found.
This excess of NBO is in agreement with recent experimental results on glasses
of composition similar to the present composition~\cite{stebbins_excess_nbo} and is
likely due to the presence of AlO$_5$ units
as well as  oxygen tri-clusters (mainly 2 Si - 1 Al).  
Also in the CAS system, it is found that the aluminum avoidance  principle is violated 
and that most of the non-bridging oxygens are located on silicon tetrahedra, 
both results being in agreement with recent experimental data 
\cite{xray2,al-avoidance1,al-avoidance2}.
Finally, evidence of the creations and
 anihilations of non-bridging 
oxygens has been obtained. These processes likely play a key role in the flow mechanism
in this system.  A detailed
understanding of the diffusion process, although 
extremely interesting, would require considerable computer time and would represent 
a methodological challenge for the {\it ab initio} approach. 
While we are currently developing new techniques to
treat this problem~\cite{Minary1}, the study of the 
dynamical properties in these systems remains, for the present, beyond the scope of this paper. 
In addition, since it is not
possible to determine which of the dominant structural motifs in CAS are 
a direct consequence of thermal fluctuations,
we intend, as future work, 
to perform a quench of the CAS system to room temperature in order to study 
its structural properties in the glassy state.


\acknowledgments

We would like to thank  Dominique Ghaleb and Jean-Marc Delaye 
who initiated the CAS simulations and
provided the classical initial sample, and Philippe Jund  
who provided  the initial  silica  sample. 
Our warm thanks  to R\'emi Jullien, Walter Kob
and Dominik Marx for very interesting and stimulating discussions and to
J\"urg Hutter for his invaluable  help with the CPMD code. 
Concerning the CAS simulations, the work of M.B. was supported by a NATO grant.
  M.E.T would like to acknowledge
support from NSF CHE-98-75824 and from the Research Corporation RI0218.
The simulations have been performed on IBM/SP2 in CINES, Montpellier, 
France and
on CRAY/T3E in IDRIS, Paris, France.



\newpage

\newpage

\begin{table}
\caption{Results of distances and angles obtained from total
 energy calculations carried out on
AlH$_2$OH, CaH$_2$ and CaH using the pseudo-potentials and 
the energy cutoffs given in the text.
A Trouiller-Martins pseudo-potential was used for hydrogen.
For these calculations, the reciprocal-space cluster boundary conditions
method of Martyna and Tuckerman was employed~\protect\cite{clus_bc}}.
\label{table:tests}
\begin{tabular}{|c|c|c|}
 {\bf AlH$_2$OH} & This work & Ref. \protect\cite{alh2oh} \\
 \hline
 Al-O (\AA) &  1.73(5) & 1.711 \\
 Al-H$_1$  (\AA) &  1.56(4) &  1.565 \\
 Al-H$_2$ (\AA) &  1.57(1) &  1.575 \\
 O-H$_3$   (\AA) & 0.97(6) & 0.956 \\
 $\widehat{\rm H_1AlH_2}$ &  125.5 &  123.9 \\
 $\widehat{\rm AlOH_3}$  & 122.3  &  122.3 \\
 \hline \hline
 {\bf CaH$_2$} & This work & Ref. \protect\cite{CaH} \\
 \hline
 Ca-H$_1$   (\AA) &   2.03(0) &  2.00 \\
 Ca-H$_2$    (\AA) &  2.03(1) &  2.00 \\
 \hline \hline
 {\bf CaH} & This work & Ref. \protect\cite{CaH} \\
 \hline
 Ca-H       (\AA) &      1.97(5) & 2.00 \\
 \end{tabular}
 \end{table}

\end{document}